\documentclass[a4paper,11pt]{article}
\pdfoutput=1 

\usepackage{jheppub} 

\usepackage{color}
\input{colordvi.tex}
\usepackage{graphicx}
\usepackage{float}
\usepackage{wrapfig}
\usepackage{bm,changes}
\usepackage{amsmath,amssymb,ascmac}
\usepackage{subfigure}
\usepackage{verbatim}
\usepackage{makeidx}
\usepackage{tikz-feynman}
\tikzfeynmanset{compat=1.0.0} 

\title{Constraining Millicharged dark matter with Gravitational positivity bounds}

 \author{Suro Kim,}
 \author{Pyungwon Ko}
 \affiliation{Korea Institute for Advanced Study, Seoul 02455, Republic of Korea}

\emailAdd{surokim@kias.re.kr}
\emailAdd{pko@kias.re.kr}

\abstract{Gravitational positivity bounds provide consistency conditions for effective field theories with gravity. They turn out to be  phenomenologically useful by  providing lower bounds in parameters of new physics beyond the Standard Models (BSM). In this paper, we derive constraints on millicharged fermion dark matter models with massless dark photon using gravitational positivity bounds. Combining them with upper bounds from cosmological and astrophysical observations, we can severely constrain the parameter space of the model. In particular, we show that when the dark matter mass is lighter than the solar core temperature, the whole parameter region is excluded by combining gravitational positivity bounds and the stellar bounds. 
}
\begin{document} 
\maketitle
\flushbottom

\section{Introduction}
\label{}

Dark matter (DM) is now the universal paradigm supported by various cosmological and astrophysical observations based on gravitational interaction. However, its extremely tiny interaction with the standard model (SM) particles makes it difficult 
to detect directly in a non-gravitational manner. The worst possibility is that DM interacts with the SM sector only through gravitational interactions, meaning ``{\it the dark matter is totally dark}".

\medskip
Considering that the standard model sector consists of many stable or long lived particles, it is rather unnatural to consider a single particle as DM compared to a dark sector. The most straightforward possibility as a dark sector would be DM which is charged under the hidden, unbroken $U(1)$ gauge symmetry. In this case, the dark photon $\gamma^\prime$ is massless and it can have interactions with the SM sector through a small kinetic mixing. Then, DM will have a tiny interaction with the SM sector and have a tiny electromagnetic charge
as well. Such a dark matter model is called as a millicharged DM and has been studied extensively
~\cite{Holdom:1985ag,Goldberg:1986nk,DeRujula:1989fe,Brahm:1989jh,Dimopoulos:1989hk,Davidson:1993sj,Davidson:2000hf,Dubovsky:2003yn,Ackerman:2008kmp,Chuzhoy:2008zy,Feng:2008mu,Gardner:2009et,Dai:2009hx,Feng:2009mn,McDermott:2010pa,Vogel:2013raa,Shiu:2013wxa,Haas:2014dda,Ejlli:2017uli,Berlin:2018bsc,Gninenko:2018ter,Kelly:2018brz,Harnik:2019zee,Liang:2019zkb,Foroughi-Abari:2020qar,Budker:2021quh,Berlin:2023gvx,Hebecker:2023qwl,Fiorillo:2024upk}.

\medskip
As a complementary method to experimental observations, it has been studied to find out consistency conditions for a UV completion of new physics. One of the well-established conditions is the positivity constraints on scattering amplitudes~\cite{Pham:1985cr,Pennington:1994kc,Ananthanarayan:1994hf,Comellas:1995hq,Adams:2006sv,Adams:2008hp}, see a review article~\cite{deRham:2022hpx}, and their phenomenological applications~\cite{Vecchi:2007na,Zhang:2018shp,Bi:2019phv,Remmen:2019cyz,Kim:2019wjo,Remmen:2020vts,Yamashita:2020gtt,Bonnefoy:2020yee,Chala:2021wpj,Li:2022rag,Kim:2023pwf,Hong:2023zgm,Kim:2023bbs}. Recently, this line of 
strategy has been extended by incorporating gravity effects to the positivity constraints~\cite{Hamada:2018dde,Bellazzini:2019xts,Loges:2020trf,Alberte:2020jsk,Tokuda:2020mlf,Herrero-Valea:2020wxz,Caron-Huot:2021rmr,Caron-Huot:2022ugt,Herrero-Valea:2022lfd,deRham:2022gfe,Noumi:2022wwf,Caron-Huot:2024tsk} and applying them to various phenomenologies~\cite{Alberte:2020bdz,Aoki:2021ckh,Noumi:2021uuv,Alberte:2021dnj,Noumi:2022zht,Hamada:2023cyt,Aoki:2023khq}.

\medskip
Let us comment on a few interesting points in phenomenological applications of the gravitational positivity bounds. First of all, constraints on renormalizable couplings can be obtained through the positivity constraints on loop-level scattering amplitudes, whereas the ordinary positivity bounds usually gives constraints on couplings whose mass dimension is equal or larger than $8$. The second point is that we can obtain lower bounds on couplings from gravitational positivity bounds. 
By combining the constraints from the gravitational positivity bounds with phenomenological upper bounds from observations, we can constrain a broad range of parameter spaces and can even exclude entire parameter space in some models. One example is the application on the massive dark photon. 
In Refs.~\cite{Noumi:2022zht,Aoki:2023khq}, the lower bound on the kinetic mixing between the photon and 
dark photon is obtained, which can be interpreted as ``{\it the dark photon cannot be too dark}". 
Also, most of the parameter region is excluded by combining the lower bounds from gravitational 
positivity bounds and phenomenological upper bounds from various observations. 
In this paper, we will elaborate on the gravitational positivity bounds in
the millicharged DM models with a massless dark photon $\gamma^\prime$ coupled to massive fermion DM.  
$\gamma^\prime$ will couple to electrically charged particles of the SM through the $U(1)$ gauge kinetic mixing.

\medskip
The organization of this paper is as follows. In Sec.~\ref{model} we will introduce the model 
for the millicharged DM and evaluate 1-loop amplitudes for 
$\gamma\gamma^\prime\to\gamma\gamma^\prime$ and 
$\gamma^\prime\gamma^\prime\to\gamma^\prime\gamma^\prime$ scatterings, where $\gamma$ and 
$\gamma^\prime$ are the photon and the dark photon, respectively. Then, in Sec.~\ref{Grav_pos} we will briefly review the gravitational positivity bounds. 
In Sec~\ref{PB_millichargedDM}, we will give constraints on parameters of the model using the gravitational positivity bounds and combine with phenomenological upper bounds obtained from 
various astrophysical and cosmological observations based, and then we will summarize.

\section{Millicharged Dark Matter}
\label{model}
\subsection{Milli-charged Dark Matter Model}
We first summarize the milli-charged dark matter model. 
We consider $U(1)_{\rm em} \times U(1)_{\rm dark}$, whose associated 
gauge fields are $A_a$ and $A_b$, respectively. 
The action of the dark sector is given by
\begin{align}
S_{\rm DS}=\int d^4x\sqrt{-g}\bigg[-\frac{1}{4}F_{b,\mu\nu}F_b^{\mu\nu}-\frac{\varepsilon}{2}F_{a,\mu\nu}F_b^{\mu\nu}+\bar{\chi}(i\gamma^\mu\nabla_\mu+e^\prime \gamma^\mu A_{b,\mu}-m)\chi\bigg]\,.
\end{align}
$F_{a,\mu\nu}$ and $F_{b,\mu\nu}$ are the field strength for each gauge boson. $\varepsilon$ measures the $U(1)$ gauge kinetic mixing and we assume that $\varepsilon\ll1$ throughout this paper. $\chi$ is a fermionic dark matter with the dark $U(1)_{\rm dark}$ charge, $e^\prime$, and mass, $m$. $\nabla$ is the covariant derivative with the spin connection. Let us perform the following field redefinition and diagonalize kinetic terms as\footnote{Note that we have one free parameter $\theta$ when we diagonalize kinetic terms~\cite{Fabbrichesi:2020wbt}. However, physical observables are independent of $\theta$ if dark photon is massless (see Ref. Xiao-Gang He et al.). The difference between basis choice would be the choice of initial and final states of scattering amplitudes to get the optimized bounds on the model parameters.}
\begin{align}
\begin{pmatrix}
A_a
\\
A_b
\end{pmatrix}
\approx
\begin{pmatrix}
1 && 0
 \\
 -\frac{\varepsilon}{\sqrt{1-\varepsilon^2}} &&\frac{1}{\sqrt{1-\varepsilon^2}}  
\end{pmatrix}
\begin{pmatrix}
\cos\theta && -\sin\theta
 \\
 \sin\theta &&\cos\theta  
\end{pmatrix}
\begin{pmatrix}
A
\\
A^\prime
\end{pmatrix}\Bigg|_{\theta=0}
+\mathcal{O}(\varepsilon^2)\,,
\label{Field_Redef}
\end{align}
up to the leading order of $\varepsilon$.

We call $A$ and $A^\prime$ photon and dark photon, respectively. Note that the diagonalization 
with $\theta=0$ generates tiny couplings 
between the photon and dark matter $\chi$, 
thereby motivating the ``milli-charged DM model":
\begin{align}
S_{\rm DS}\approx\int d^4x\sqrt{-g}\bigg[&-\frac{1}{4}F^\prime_{\mu\nu}F^{\prime\,\mu\nu}+\bar{\chi}\left(i\gamma^\mu\nabla_\mu+e^\prime \gamma^\mu A^\prime_{\mu}-\frac{\varepsilon}{2} e^\prime\gamma^\mu A_\mu-m\right)\chi+\mathcal{O}(\epsilon^2)\bigg]\,.
\end{align} 
In the next subsection, we will analyze $\gamma^\prime \gamma^\prime \to \gamma^\prime \gamma^\prime$ 
and $\gamma \,\gamma^\prime \to \gamma\, \gamma^\prime$ scatterings 
through the loops involving DM fermions and electrically charged SM particles.

\subsection{Scattering amplitudes}
\begin{figure}[t] 
	\centering 
	\includegraphics[width=4.5cm]{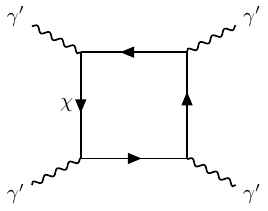}	\quad
	\includegraphics[width=4.5cm]{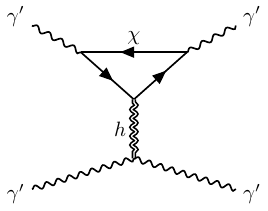}
	\caption{Feynman diagrams for $\gamma^\prime \gamma^\prime \to \gamma^\prime \gamma^\prime$.} \label{FeynD1}
\end{figure}
Let us define the following helicity combinations of the scattering amplitudes for 
$\gamma^\prime \gamma^\prime \to \gamma^\prime \gamma^\prime$ and 
$\gamma\gamma^\prime\to \gamma\gamma^\prime$: 

\medskip
\begin{align}
\mathcal{M}=\frac{1}{4}\big[ \mathcal{M}(1^+2^+3^+4^+)+\mathcal{M}(1^+2^-3^+4^-)+\mathcal{M}(1^-2^+3^-4^+)+\mathcal{M}(1^-2^-3^-4^-)\big] \,.
\end{align}
The superscript $\pm$ represents the helicity of each external particle.
Note that we only have transverse modes because both photon and dark photon are massless in our model.

\medskip
Let us first evaluate $\gamma^\prime \gamma^\prime \to \gamma^\prime \gamma^\prime$.
We use the Mathematica packages {\sc FeynArts}~\cite{Hahn:2000kx} and {\sc FeynCalc}~\cite{Shtabovenko:2023idz}, to calculate the one-loop diagrams and {\sc Package-X}~\cite{Patel:2016fam} 
to evaluate the loop integrals.
The leading non-gravitational contribution comes from the dark matter loop (the left figure in  Fig.~\ref{FeynD1}), which is given by
\begin{align}
\nonumber
\mathcal{M}_{{\rm QED}^\prime}=&-\frac{2\alpha^{\prime\,2}}{s^2}\Bigg[2\sqrt{s(s-4m^2)}(s+4m^2)\log\left(\frac{\sqrt{s(s-4m^2)}+2m^2-s}{2m^2}\right)
\\\nonumber
&\qquad\quad+(s^2+4m^2s-8m^4)\log^2\left(\frac{\sqrt{s(s-4m^2)}+2m^2-s}{2m^2}\right)+6s^2\Bigg]
\\&+(s\leftrightarrow-s)+\mathcal{O}(t)\,.
\label{Amp_4-prime_QED}
\end{align}
where $\alpha^\prime=e^{\prime\,2}/4\pi$. We expanded the amplitude around $t=0$ for later 
convenience. 

On the other hand, the contribution from the graviton exchange 
(the right figure in Fig.~\ref{FeynD1}) is given by
\begin{align}
\mathcal{M}_{\rm GR}= 
&- \frac{\alpha^\prime}{360\pi m^2 M_{\rm Pl}^2 s}
\Bigg[120m^2\sqrt{s(s-4m^2)}(s+5m^2)\log\left(\frac{\sqrt{s(s-4m^2)}+2m^2-s}{2m^2}\right)
\nonumber\\
&\qquad\qquad\qquad\qquad+180m^4(s+2m^2s)\log^2\left(\frac{\sqrt{s(s-4m^2)}+2m^2-s}{2m^2}\right)
\nonumber\\
&\qquad\qquad\qquad\qquad+s(11s^2+ 1560 m^4)\Bigg]+(s\leftrightarrow-s)+\mathcal{O}(t)\,,
\label{4-prime_grav_amp}
\end{align}
in harmonic/de Donder gauge. Note that we have already performed the wave function renomalization of external dark photons.

\begin{figure}[t] 
	\centering 
	\includegraphics[width=4.5cm]{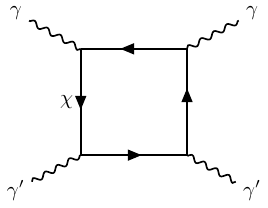}	\quad
	\includegraphics[width=4.5cm]{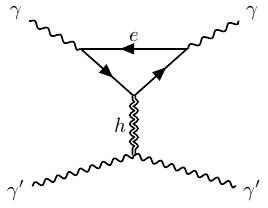}	\quad
	\includegraphics[width=4.5cm]{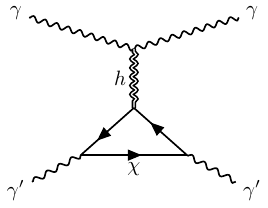}
	\caption{Feynman diagrams for $\gamma \gamma^\prime \to \gamma \gamma^\prime$ mediated by the electron and the dark matter.} \label{FeynD2}
\end{figure}
\medskip

Next, let us move on to $\gamma\gamma^\prime\to \gamma\gamma^\prime$ scattering. Without graviton exchange, we have contributions from DM loop contribution which has the same form as~\eqref{Amp_4-prime_QED} with an additional factor $\epsilon^2$. 
The gravitational contribution to the $\gamma\gamma^\prime\to \gamma\gamma^\prime$ scattering 
around $t=0$
(the middle and the right Feynman diagrams of Fig.~\ref{FeynD2}) is given by
\begin{align}
\mathcal{M}_{\rm GR}=  -\frac{11\alpha s^2}{360\pi m_e^2 M_{\rm Pl}^2}-\frac{11\alpha^\prime s^2}{360\pi m^2 M_{\rm Pl}^2}+\mathcal{O}(t)\,,
\label{2-prime_grav}
\end{align}
after wave function renormalization of external particles. 

\section{Brief Review on Gravitational Positivity Bounds}
\label{Grav_pos}
In this section, we introduce observables which we will calculate in the next section and the consequences from the gravitational positivity bounds. Throughout this paper, we will follow the notation and the workflow well established in~\cite{Aoki:2023khq} .

\medskip
Let  $\mathcal{M}(s,t)$ be the $s$-$u$ symmetric scattering amplitude for $AB\to AB$, 
where $s$, $t$, and $u$ are the Mandelstam variables. 
We assume the following properties for $\mathcal{M}$. 
\begin{enumerate}
\item{Analyticity:}
On the physical sheets of complex $s$-plane, $\mathcal{M}(s,t\to-0)$ is analytic, excluding poles and branch cuts on the real axis of $s$, which correspond to the on-shell tree-level and loop contributions to $\mathcal{M}$, respectively. 
\item{Unitarity:}
The imaginary part of $\mathcal{M}$ is non-negative. 
\item{Mild behavior at UV:}
$\mathcal{M}$ is bounded by $s^2$ in the high energy, 
\begin{align}
\lim_{|s|\to \infty}\frac{\mathcal{M}(s,t\to-0)}{s^2}=0\,.
\end{align}
\end{enumerate}

\medskip
Let us elaborate more on the last assumption. In the absence of gravity, it is guaranteed in local QFTs 
by the Froissart-Martin bound~\cite{Froissart:1961ux,Martin:1962rt,Azimov:2011nk}. On the other hand, it has some subtleties when we incorporate gravity into the discussion. Scattering amplitudes for graviton 
exchanges in the $t$-channel include terms such as $\sim -s^2/(M_{\rm Pl}^2 t)$, which breaks the third assumption for the fixed negative $t$.
To address the issue, in~\cite{Tokuda:2020mlf} the authors suggested to assume the Regge behavior of high energy scatterings:
\begin{align}
\lim_{|s|\to \infty}  {\rm Im} \mathcal{M}(s,t)\approx f(t)\left(\frac{\alpha^\prime s}{4}\right)^{2+j(t)}\,,
\end{align}
which is realized in a weakly coupled UV completion of gravity. $j(t)$ is negative for fixed negative $t$ so that it satisfies the third assumption. $\alpha^\prime$ is the scale of Reggeization. Having such cases in mind, we will assume that the third condition holds, even though we will not specify the UV theory throughout this paper.

\medskip
Next, we briefly review what the gravitational positivity bounds states.\footnote{See~\cite{Tokuda:2020mlf} for the detailed derivation. Also refer to~\cite{Aoki:2023khq} for the guidelines for application to phenomenology.} In this paper, we focus on the case that the external particles are all massless so that $s+t+u=0$. We expand $\mathcal{M}$ in terms of
the Mandelstam variables in the low energy limit. Let $a_2$ be 
an $s^2$ coefficient of a scattering amplitude after subtracting the graviton $t$-channel pole;
\begin{align}
a_2:=\lim_{t\to-0}\left[\frac{\partial^2\mathcal{M}(s,t)}{\partial s^2}+\frac{2}{M_{\rm Pl}^2t} \right]_{s=0}\,.
\label{a2}
\end{align}
The first term can be rewritten by picking up the residue at the origin ($s=0$) ,
\begin{align}
a_2=\lim_{t\to-0}\left[\frac{1}{\pi i}\oint ds\frac{\mathcal{M}(s,t)}{s^3}+\frac{2}{M_{\rm Pl}^2t} \right]_{s=0}\,. 
\end{align}
From the assumption 1 and 3, we can justify the following deformation 
of the integration contour for the first term,
\begin{align}
a_2=\lim_{t\to-0}\left[\frac{4}{\pi}\int_{m_{\rm th}^2}^\infty ds\frac{{\rm Im} \mathcal{M}(s,t)}{s^3}+\frac{2}{M_{\rm Pl}^2t} \right]_{s=0}\,,
\end{align}
where $m_{\rm th}$ is the mass of the lightest intermediate state. In our case, $m_{\rm th}^2=4m_e^2$ when $m_e<m$ or $m_{\rm th}^2=4m^2$ when $m<m_e$.
Next, we introduce the reference scale of the theory, $\Lambda$, below which we assume our EFT is valid. 
Below $\Lambda$, $\mathcal{M}$ is calculable quantity within our framework, then we can define $B(\Lambda)$ as a $s^2$ coefficient subtracting contributions below $\Lambda$:
\begin{align}
B(\Lambda)&:=a_2-\frac{4}{\pi}\int_{m_{\rm th}^2}^{\Lambda^2}ds\frac{{\rm Im} \mathcal{M}(s,t=0)}{s^3}\,.
\label{b2}
\end{align}
The gravitational positivity bounds states that there exist the lower bound on $B(\Lambda)$~\cite{Tokuda:2020mlf}
\begin{align}
B(\Lambda)\gtrsim \frac{\sigma}{M_{\rm Pl}^2 M^2}\,,
\label{positivity_bound}
\end{align}
The sign $\sigma=\pm 1$, and the scale $M$ are determined by details of UV behaviors of scattering amplitudes, e.g. the Regge behavior. Throughout this paper, we will assume that $M$ is larger than $\Lambda$, the electron mass $m_e$ and a dark matter mass $m$. In this case, we can ignore the r.h.s of~\eqref{positivity_bound}, then the bound is simplified as
\begin{align}
B(\Lambda)\gtrsim 0\,.
\label{grav_posit_prac}
\end{align}

\medskip
Since we include the gravity in discussion, $B(\Lambda)$ contains both gravitational contributions $B_{\rm grav}(\Lambda)$, and non-gravitational contributions $B_{\rm non-grav}(\Lambda)$. Among them, $B_{\rm non-grav}(\Lambda)$ give always positive contributions whereas $B_{\rm grav}(\Lambda)$ might give negative contributions as we will see later. Therefore, non-gravitational contribution should compensate the gravitational contribution to satisfy the bound~\eqref{grav_posit_prac}. Such behavior implies that the gravitational positivity bounds constrain renomalizable couplings, whereas the non-gravitational positivity bounds usually constrain nonrenormalizable couplings such as dimension-8 couplings.

\section{Gravitational positivity bounds on millicharged DM}
\label{PB_millichargedDM}
In this section, we obtain the constraints from the gravitational positivity bounds on the 
millicharged DM model and compare with the phenomenological constraints from astrophysical and  
cosmological observations.

\subsection{Constraints from $\gamma^\prime\gamma^\prime\to\gamma^\prime\gamma^\prime$ scattering}
Let us start from the bound related to  $\gamma^\prime\gamma^\prime\to\gamma^\prime\gamma^\prime$ scattering. To get a bound on the model parameters, $\alpha^\prime$, $\varepsilon$, and $m$, we evaluate $B(\Lambda)$ for each scattering amplitude.
Using Eq.~\eqref{a2} with the dark matter loop contribution~\eqref{Amp_4-prime_QED}, we can obtain $B(\Lambda)_{{\rm QED}^\prime}$ as
\begin{align}
&B_{{\rm QED}^\prime}\approx\frac{16\alpha^{\prime\,2}}{\Lambda^4}\left(\ln\frac{\Lambda}{m}-\frac{1}{4}\right)\,,
\label{4-prime_QED}
\end{align}
where $\Lambda\gg m_e$.
For the gravitational contribution, only the $t$-channel diagram contribute to $B(\Lambda)$. They don't have the $s$-channel and $u$-channel discontinuities so that we can evaluate $B(\Lambda)$ as a $a_2$
\begin{align}
&B_{\rm GR}\approx a_{2,{\rm grav},t-{\rm channel}}=-\frac{11\alpha^\prime}{90\pi m^2M_{\rm Pl}^2}\,,
\label{4-prime_grav}
\end{align}
Note that the gravitational contribution is negative so that we can obtain non-trivial constraints. Since~\eqref{4-prime_grav} is dominant over the r.h.s of~\eqref{positivity_bound} when $M\gg m/{e^\prime}$, we can use~\eqref{grav_posit_prac}. As a result, we obtain the following lower bound on $\alpha^\prime$ : 
\begin{align}
\alpha^\prime\left(\ln\frac{\Lambda}{m}-\frac{1}{4}\right)>\frac{11}{1440\pi}\left(\frac{\Lambda}{m}\right)^2\left(\frac{\Lambda}{M_{\rm Pl}}\right)^2\,.
\label{m-a_grav}
\end{align}
\begin{figure}[t]
	\centering 
	\includegraphics[width=7.5cm]{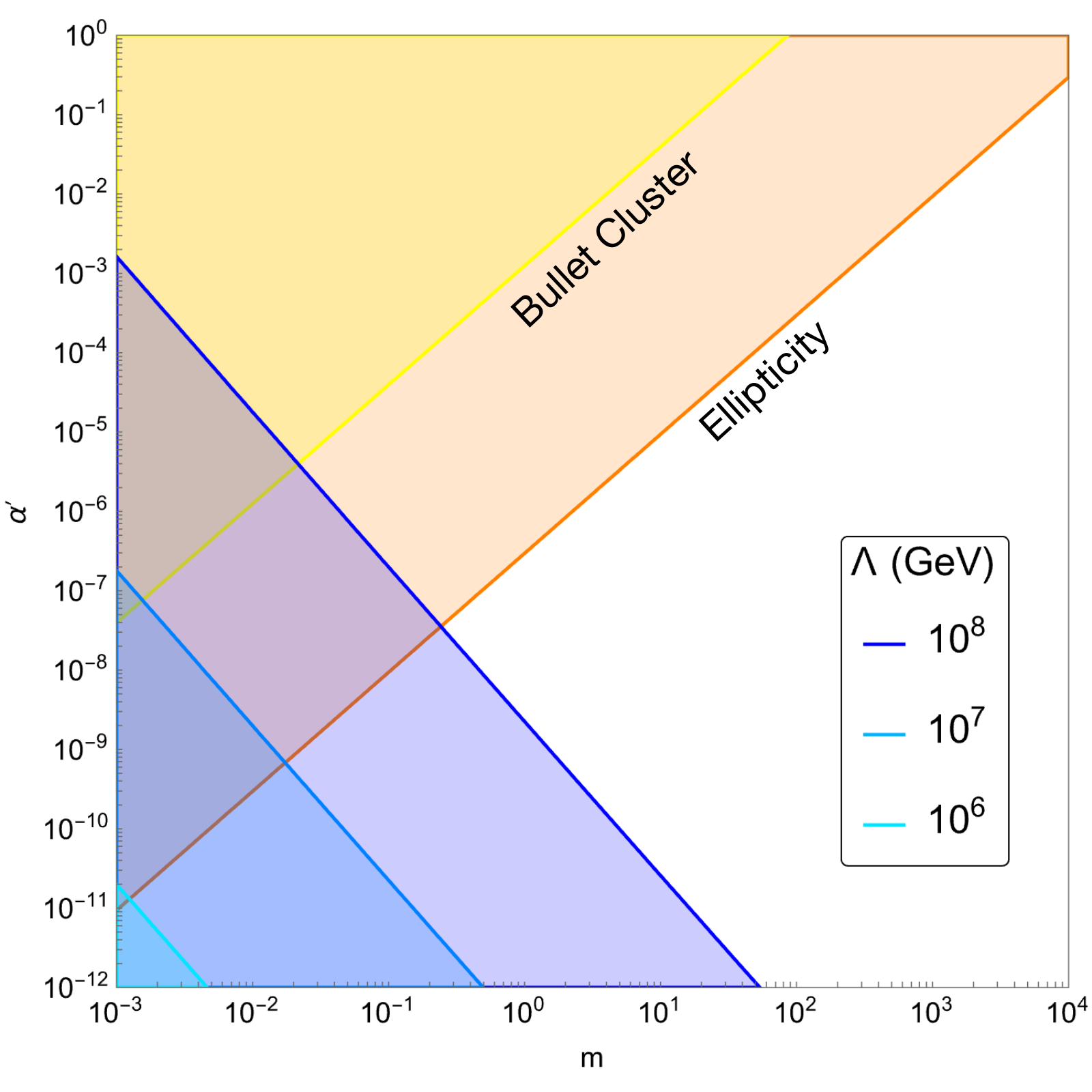}
	\caption{Constraints on $m$-$\alpha^\prime$ plane from the gravitational positivity bounds and cosmological constraints. The blue regions are excluded by the gravitational positivity bounds for $\Lambda=10^8$, $10^7$, and $10^6$ GeV, respectively. The orange region is constrained by ellipticity of galactic dark matter halos and the yellow region is constrained by the Bullet Cluster~\cite{Feng:2009mn}. } \label{M-a}
\end{figure}

This bound is plotted in Fig.~\ref{M-a} with the corresponding cosmological and 
astrophysical constraints~\cite{Feng:2009mn}. The DM self-interaction mediated by the dark photon is constrained by the Bullet Cluster and an ellipticity of galactic dark matter halos, as illustrated with the yellow and orange region, respectively.  
Note that that the higher $\Lambda$ leads to the stronger gravitational positivity bounds. 

\medskip
One can check that in the case of the light dark matter, $m\sim10^{-3}$ GeV and when $\Lambda\gtrsim10^6$ GeV, the whole region of $\alpha^\prime$ is excluded by combining the cosmological and astrophysical constraints and the gravitational positivity bounds. On the other hand, when the dark matter mass becomes larger, both constraints become weaker and a broad region is allowed.

\subsection{Constraints from $\gamma\gamma^\prime\to\gamma\gamma^\prime$ scattering}

\begin{figure}[t] 
	\centering 
	\includegraphics[width=8.5cm]{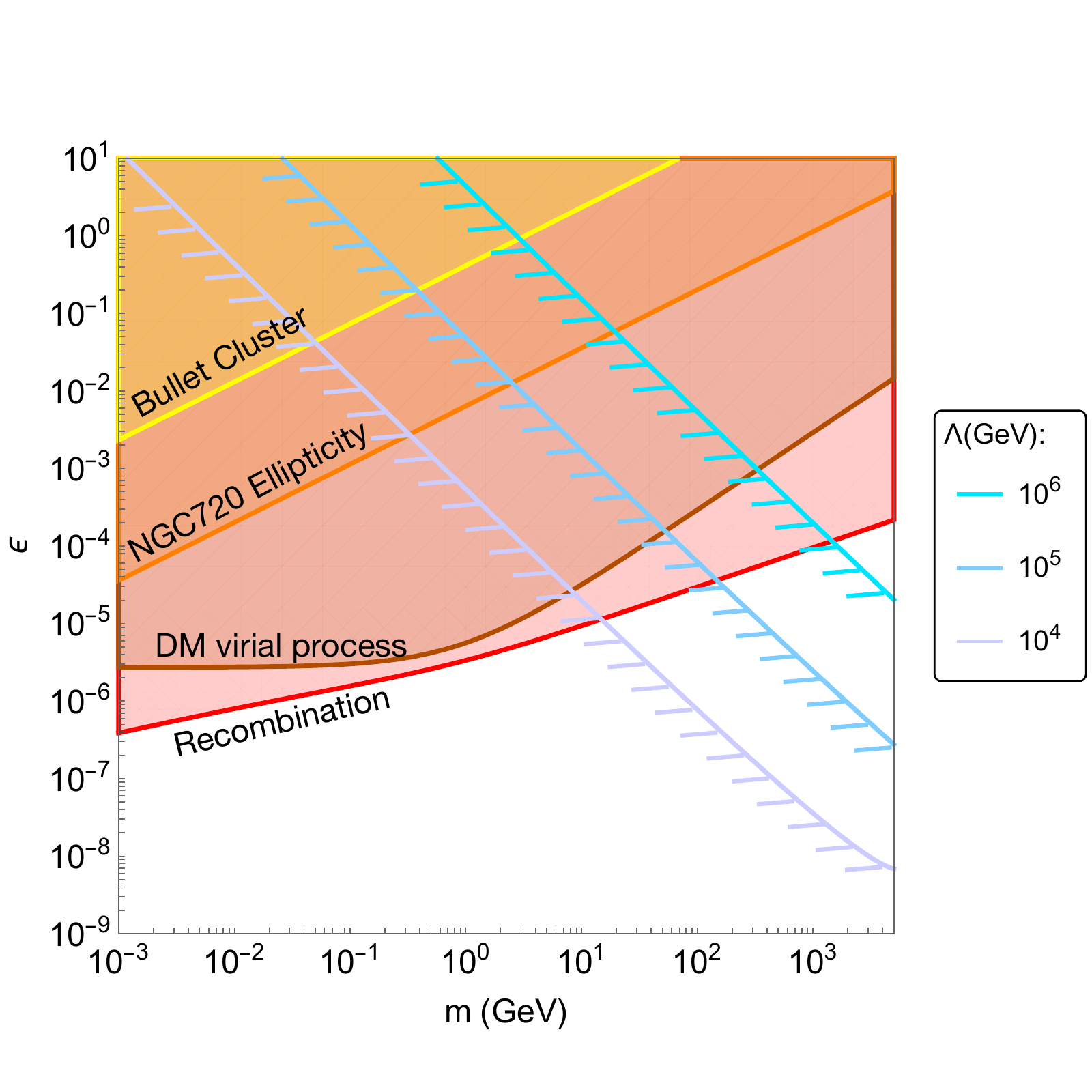}
	\caption{Constraints on $m$-$\varepsilon$ plane from the gravitational positivity bounds and cosmological constraints. Considering that the smaller $\alpha^\prime$ leads to the stronger bound, we assumed the biggest value for $\alpha^\prime$ allowed from ellipticity of galactic dark matter halos~\cite{Feng:2009mn}.  The regions below the each blue lines are excluded by the gravitational positivity bounds for $\Lambda=10^6$, $10^5$, and $10^4$ GeV, respectively. The behavior of the blue lines mainly comes from the weakened upper bound of $\alpha^\prime$ for heavier dark matter mass (see Fig.~\ref{M-a}). The red, brown, orange, and yellow regions are constrained by decoupling before the recombination era, dark matter virial process, of galactic dark matter halos, 
and the Bullet Cluster, respectively.} \label{M-e}
\end{figure}

\medskip
Next, we move on to the bounds from $\gamma\gamma^\prime\to \gamma\gamma^\prime$ scattering. 
In the same way as the previous subsection,, we can evaluate $B(\Lambda)$ for the dark matter loop contribution and gravitaton $t$-channel exchange:
\begin{align}
&B_{\chi}\approx\frac{16\varepsilon^2\alpha^{\prime\,2}}{\Lambda^4}\left(\ln\frac{\Lambda}{m}-\frac{1}{4}\right)\,.
\label{2-prime_QED}
\\
&B_{\rm GR}\approx-\frac{11\alpha}{180\pi m_e^2M_{\rm Pl}^2}-\frac{11\alpha^\prime}{180\pi m^2M_{\rm Pl}^2}\,,
\label{2-prime_grav}
\end{align}
where we have taken $\Lambda\gg m$.

\medskip
Similarly to the previous subsection, when $M\gg m_e/e$, $m/{e^\prime}$, 
the r.h.s of~\eqref{positivity_bound} can be ignored so that we can use~\eqref{grav_posit_prac} to derive the following bounds: 
\begin{align}
\varepsilon^2\alpha^{\prime\,2}\left(\ln\frac{\Lambda}{m}-\frac{1}{4}\right)>\frac{11\alpha}{2880\pi}\left(\frac{\Lambda}{M_{\rm Pl}}\right)^2\left(\frac{\Lambda}{M_{\rm e}}\right)^2\left[1+\frac{\alpha^{\prime}}{\alpha}\left(\frac{m_e}{m}\right)^2\right]\,.
\label{epsilon-alpha_positivity}
\end{align}

\medskip
Let us first focus on the WIMP mass range, $10^{-3}$ GeV $\lesssim m \lesssim 10^{4}$ GeV. See Fig.~\ref{M-e} for the positivity bounds and cosmological and astrophysical constraints in
the $m-\varepsilon$ plane. Note that kinetic mixing $\varepsilon$ generates DM couplings to SM sector, that gives following cosmological constraints~\cite{McDermott:2010pa}. DM should decouple from the SM sector before recombination, not to affect to the CMB data, whose bound on $\varepsilon$ is depicted by the red region in Fig.~\ref{M-e}. On top of that, if DM couples too strongly to the SM sector after recombination, the SM sector transfers energy to DM and it modifies the DM virialization process, whose bound is shown as the brown region in Fig.~\ref{M-e}. On the other hand, $\varepsilon$ also generates the DM self-interaction through the photon exchange. It gives constraints from the Bullet cluster and ellipticity of galactic dark matter halos, same as the case of $\alpha^\prime$, which are shown as the yellow and orange region in Fig.~\ref{M-e}, respectively. Note that a broad range of the parameter space is excluded by combining constraints from observations and the gravitational positivity bounds. 

\medskip
We can interpret the result as follows. Suppose we observed the millicharged dark matter whose mass is smaller than $10$ GeV. Then, the dark sector should be modified under the $\Lambda\sim10^4$ GeV. For example, we can extend the dark gauge structure by incorporating dark electroweak sector or dark QCD sector. In~\cite{Aoki:2021ckh}, the authors studied the application of the gravitational positivity bounds to the Standard model. If we consider QED coupled to GR, the gravitational positivity bounds give the cutoff scale $\Lambda_{\rm QED}\sim10^8$GeV. If we include the weak sector, the cutoff scale can be raised to $\Lambda_{\rm EW}\sim3.8\times 10^{13}$ GeV. On the top of that, we can obtain an even larger cutoff scale $\Lambda_{\rm SM}\sim3\times 10^{16}$ GeV by incorporating QCD contributions. On the other hand, if we do not observe any clue for the millicharged dark matter from the future experiments, the observational constraints will become stringent. Then, we can exclude the millicharged dark matter model by combining experimental constraints and the gravitational positivity bounds we considered in this paper.

\begin{figure}[t] 
	\centering 
	\includegraphics[width=7cm]{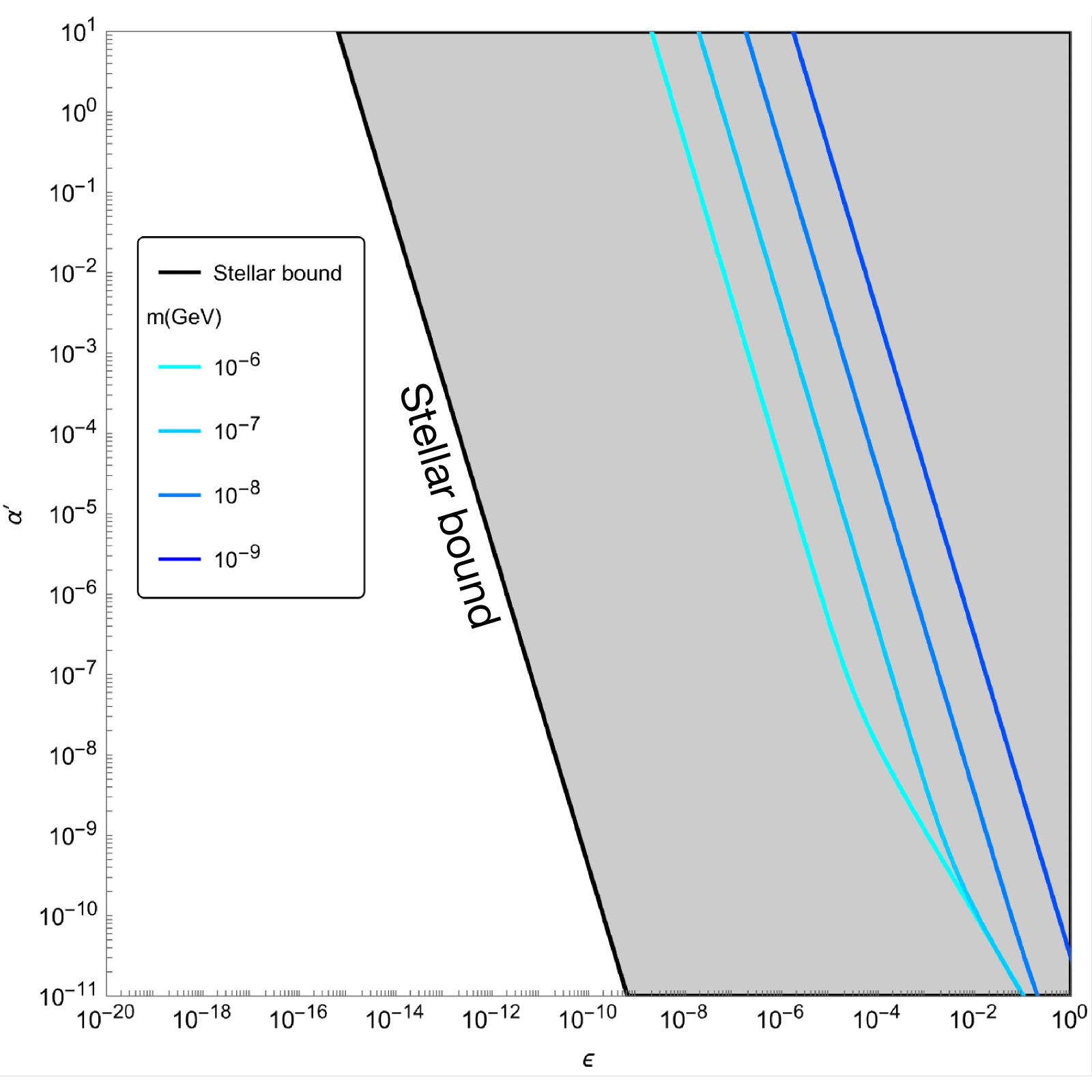}\qquad
	\includegraphics[width=7cm]{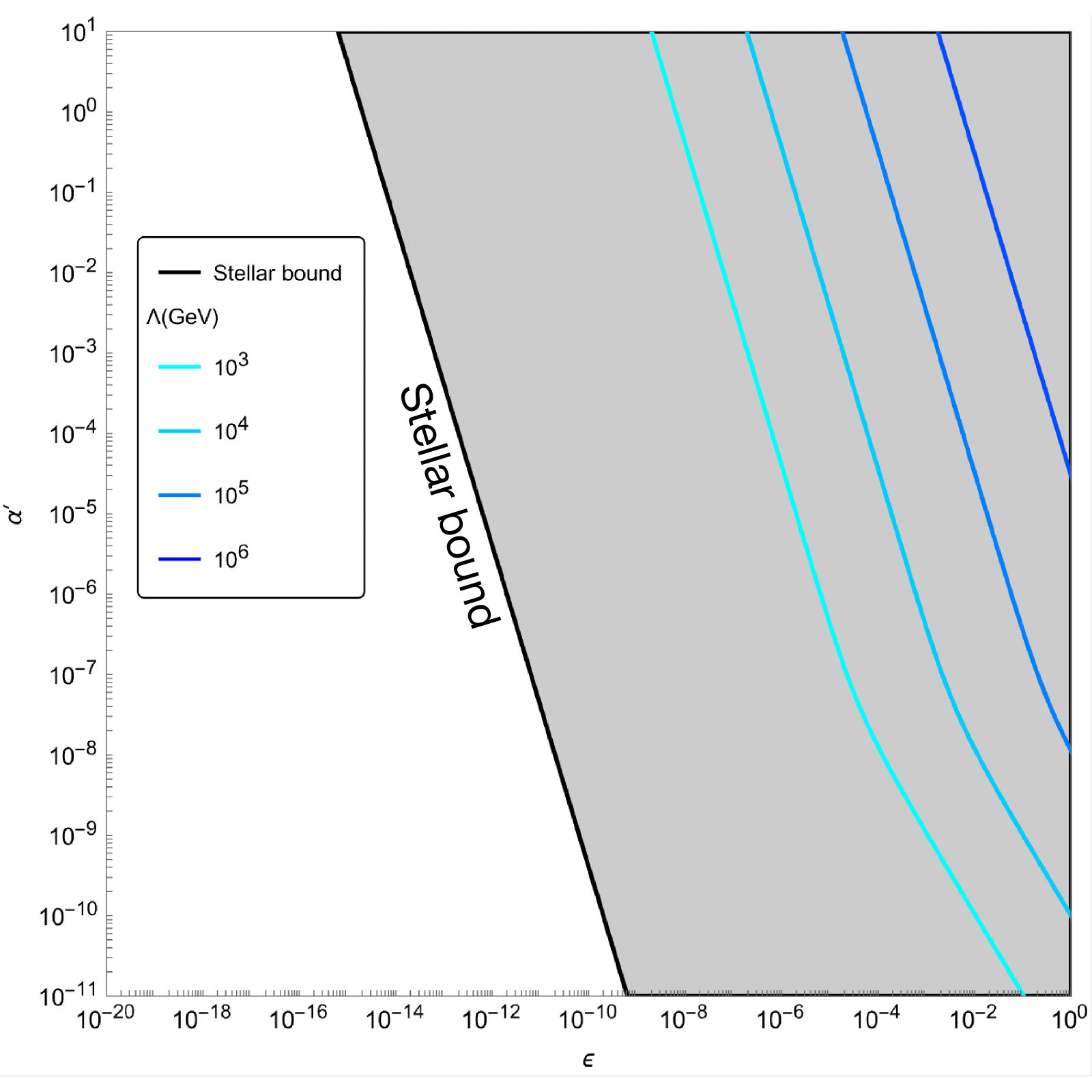}
	\caption{The constraints on $\varepsilon$-$\alpha^\prime$ plane from the gravitational positivity bounds and the stellar bound. The shaded region is forbidden by the stellar bound. The left region of each blue line is excluded. In the left plot, we fixed the reference scale $\Lambda$ and vary DM mass $m$, whereas in the right plot we fixed $m$ and vary $\Lambda$. (Left) Each blue lines are for $m=10^{-6}$, $10^{-7}$, $10^{-8}$, and $10^{-9}$ GeV, where $\Lambda=10^3$ GeV. (Right) Each blue lines are for $\Lambda=10^{3}$, $10^{4}$, $10^{5}$, and $10^{6}$ GeV, where $m=10^{-6}$ GeV.} \label{e-a}
\end{figure}

\medskip
Next, we move on to the lighter dark matter case, whose mass is lighter than the solar core temperature, $T_\odot\sim 1$ keV. In this mass region, there exist various phenomena such as dark solar wind~\cite{Chang:2022gcs}. The most relevant cosmological constraint for this case comes from the stellar bound, which states that the coupling between a light dark matter and the SM particles should be small enough to avoid a pair creation of DM from the SM plasma inside the Sun and anomalous evolution in red giants. The most stringent stellar bound is given by~\cite{Vogel:2013raa}
\begin{align}
\varepsilon \alpha^{\prime\,1/2}\lesssim2\times 10^{-15}\,.
\end{align}
See Fig.~\ref{e-a} for the constraints on $\varepsilon$-$\alpha^\prime$ plane. In the left panel, we fixed $\Lambda=1$ TeV and varied $m$, whereas on the right panel we fixed $m=10^{-6}$ GeV and varied $\Lambda$. We took $\Lambda\gtrsim 1$ TeV, assuming no dark particles coupled to the photon are found below $1$ TeV.

\medskip
Note that the larger $\Lambda$ and the lower $m$, the stronger bound from gravitational positivity. 
By combining the stellar bound~\cite{Vogel:2013raa} and the gravitational positivity bounds, we can  exclude the whole parameter region.\footnote{In~\cite{Abu-Ajamieh:2024gaw}, a qualitatively similar bound is obtained from the other consistency condition with quantum gravity, the weak gravity conjecture. Nevertheless, our analysis yields a quantitatively stronger bound.} Note that we didn't assume that this light millicharged particle is a dark matter.


\section{Conclusion}
In this paper, we discussed the application of the gravitational positivity bounds on the millicharged dark matter model and obtained constraints on the model parameters $\alpha^\prime$, $\varepsilon$, and $m$. In particular, in the case that the dark matter mass is below the solar core temperature $T_\odot\sim1$ keV, we could exclude the whole parameter region (see Fig.~\ref{e-a}).

\medskip
In the case of the WIMP mass range, we obtained the relatively relaxed bound, which is plotted in Fig.~\ref{M-a} and Fig.~\ref{M-e}. When the dark matter mass is light, $m<10$ GeV, and for $\Lambda>10^4$ GeV, all the region of $\alpha^\prime$ and $\varepsilon$ is excluded by combining the cosmological constraints and the gravitational positivity bounds.


\medskip
Note that the gravitational positivity bounds are applicable to various dark matter models. 
In particular, they can be highly advantageous for scenarios wherein observational data stringently restrict the parameter space. An immediate example would be non-Abelian dark gauge 
models, in which interactions between dark matter and dark radiation can affect the matter power 
spectrum and $\sigma_8$, as well as dark radiation itself can change the Hubble parameter $H_0$
\cite{Buen-Abad:2015ova,Ko:2016fcd}. 
The gauge coupling should be very tiny so that the matter power spectrum is not suppressed 
too much. Investigating whether the phenomenologically acceptable 
dark gauge coupling satisfies the gravitational positivity bounds would be interesting.
We leave such discussions for future works.

%

\acknowledgments 
We draw Feynman diagrams using the Mathematica package TikZ-Feynman~\cite{Ellis:2016jkw}.
This work is supported in part by KIAS Individual Grants under Grant No. PG082102 (S.K.), and No. PG021403 (P.K.) at Korea Institute for Advanced Study.

\appendix

\bibliography{bib.bib}{}
\bibliographystyle{utphys}

\end{document}